\documentclass[epj]{svjour}
\usepackage [dvips] {graphics}
\begin{document}
\title{STRING and PARTON PERCOLATION}
\author{C. Pajares\inst { }
}                     
%
%
\institute{Departamento de F\'{\i}sica de Part\'{\i}culas Elementais e Instituto Galego de F\'{\i}sica de Altas Enerx\'{\i}as\\ Universidade de Santiago de Compostela, 15782 Santiago de Compostela, Spain}
\date{Received: date / Revised version: date}
%
\abstract{
A brief review to string and parton percolation is presented. After a short introduction, the main consequences of percolation of color sources on the following observables in A-A collisions: $J/\psi$ suppression, saturation of the multiplicity, dependence on the centrality of the transverse momentum fluctuations, Cronin effect and transverse momentum distributions, strength of the two and three body Bose-Einstein correlations and forward-backward multiplicity correlations, are presented. The behaviour of all of them can be naturally explained by the clustering of color sources and the dependence of the fluctuations of the number of these clusters on the density.
     } 
%
\maketitle
\section{Introduction}
\label{intro}
What conditions are necessary in the pre-equilibrium stage to achieve deconfinement
and perhaps subsequent quark-gluon plasma formation? This question on the occurence of color
deconfinement in nuclear collisions without assuming prior equilibration has been addressed
on the basic of two closely related concepts, string or parton percolation \cite{Ref1}-\cite{Ref2} and parton
saturation \cite{Ref3}-\cite{Ref4}-\cite{Ref5}.In this paper we will study the first subject.

Consider a flat two dimensional surface S (the transverse nuclear area), on which $N$ small
disc of radius $r_0$ (the transverse partonic or string size) are randomly distributed, allowing
overlapping. With increasing density $n\equiv N/\pi R^{2}$ (we take here $S=\pi R^2)$, clusters
of increasing size appear. The crucial feature is that this cluster formation shows critical
behaviour: in the limit $N\rightarrow\infty$ and $R\rightarrow\infty$ with $n$ finite, the
cluster size diverges at a certain critical density. The percolation threshold is given by

\begin{equation}
	\eta_{c}=\pi r^2_0 \frac{N}{\pi R^2}
\end{equation}
and its value 1.13 is determined by numerical studies. For finite $N$ and $R$, percolation
sets in when the largest cluster spans the entire surface from the center to the edge.
Because of overlap, a considerable fraction of the surface is still empty at the percolation
point in fact, at the threshold, only $1-exp(-\eta_{c})\simeq 2/3$ of the surface is covered
by discs.

In high energy nuclear collision, the strings or partons are originated from the nucleons
within the colliding nuclei, therefore their distribution on the transverse area of the
collision is highly non uniform with more nucleons and hence more strings or partons
in the center than in the edge. In this case the value of $\eta_c$ becomes higher \cite{Ref6}.

\section{Local parton percolation and $J/\psi$ suppression}
Hard probes, such as quarkonia, probe the medium locally, and thus test only
if it has reached the percolation point and the resulting geometric deconfinement at their
location. It is thus necessary to define a more local percolation criterium \cite{Ref7}.

As we mentioned before, at the percolating critical density, 1/3 of the surface remains empty.
Hence disc density in the percolating cluster must be greater than
$\frac{3}{2}\frac{\eta_c}{\pi r^2_0}$.
In fact numerical studies show that percolation sets in when the density of partons in the 
largest cluster reaches the critical value $1.72/\pi r^2_0$, slightly larger than 
$\frac{3}{2}\frac{1.13}{\pi r^2_0}$.
This result provides the required local test: if the parton density at a certain point in
the transverse nuclear collision plane has reached this level, the medium there belongs to
a percolating cluster and hence to a deconfined parton condensate. In Fig. 1 the percolation
probability and its derivative as a function of $\eta$ are shown.

\begin{figure}
\resizebox{0.45 \textwidth}{!}{%
  \includegraphics{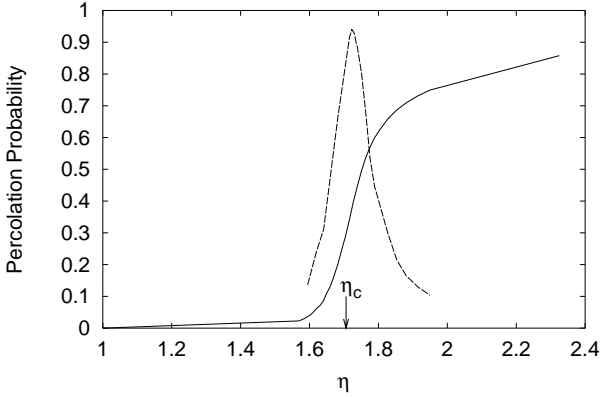}
}
\caption{Percolation probability and its derivative as a function of $\eta$}
\end{figure}

Let us apply the above idea to $J/\psi$ suppression in A-A collision \cite{Ref8}. 
We denote by $n_s (A)$ the density of nucleons in the transverse plane and by
$dN_q(x,Q^2)/dy$
the parton distribution functions. (At central rapidity $y=0$, we have
$x=k_T/\sqrt{s}$, where $k_T$ denotes the transverse momentum of the partons and thus $k_T \simeq Q$).
The local parton percolation condition is

\begin{equation}\label{eq2}
n_s(A) \left(\frac{dN_q(x,Q^2_c)}{dy}\right)_{x=Q_C/\sqrt{s}}=\frac{1.72}{\pi/Q^2_c}
\end{equation}

For a given A-A collision at a fixed centrality and energy, the relation (\ref{eq2}) determines $Q_C$.

For $Pb-Pb$ collision at $\sqrt{s} = 17.4\ GeV$, $Q_c\simeq 0.7\ GeV$.
The scales of the charmonium states $\chi_c$ and $\psi'$, as determined by the 
inverse of their radii calculated in potential theory, are around 0.6 GeV and 0.5 GeV respectively,
therefore the parton condensate can thus resolve these states and all $\chi$ and $\psi'$ states
formed inside the percolating cluster disappear. The location is
determined by the collision density. The first onset of $J/\psi$ suppression in $Pb-Pb$ collision
at SPS should occur at $N_{part}\simeq 125$, where the $J/\psi's$ due to feed-down from $\chi_c$ and 
$\psi'$ states in the percolating cluster are eliminated. Directly produced $J/\psi's$ survive
because of their smaller radii (leading to a scales of 0.9--1.0 GeV) and its dissociation
requires more central collisions, which lead to a better resolution, i.e. to an increase of $Q_c$.
For $Q_c=1.0\ GeV$ we need $N_{part}\simeq 320$. The resolution scale of the direct $J/\psi$ cannot be reached in S-U collisions, therefore only one stop pattern suppression is obtained for this case.

For Au-Au collision at RHIC, the increase parton density shift the onset of percolation to a 
higher resolution scale, so that from the threshold on, all charmonium states including $J/\psi's$ are
supressed, starting at $N_{part}\simeq 90$, i.e. a single step suppression pattern occurs.

For the case In-In collisions at SPS energies, the 
thres\-hold for directly produced $J/\psi's$ is not
reached even for the most central collisions, and again a single step suppression patterm is expected.

A detailed discussion and comparison with experimental data can be found in references \cite{Ref7} and \cite{Ref9}.

\section{String percolation}
Multiparticle production is currently described in terms of color strings stretched between
partons of the projectile and target, which decay into new strings through $q-\bar{q}$ production 
and subsequently hadronize to produce observed hadrons. Color strings may be viewed as 
small discs in the transverse space, $\pi r^2_0$, $r_0=0.2-0.25\ fm$, filled with the
color field created by the colliding partons. Particles are produced by the Schwinger
mechanisms \cite{Ref10} emitting $q\bar{q}$ pairs in this field. With growing energy and/or atomic
number of colliding particles, the number of strings grows and they start to overlap, forming clusters.
At a critical density a macroscopic cluster appears that marks the percolation phase transition.

The percolation theory governs the geometrical pattern of the string clustering. Its observable implications,
however, require introduction of some dynamics to describe string interaction, i.e, the behaviour of 
a cluster formed by several overlapping strings.

It is assumed that a cluster behaves as a single string with a higher color field
$\stackrel{\rightarrow}{Q_n}$ corresponding to the vectorial sum of the color charge of each individual
$\stackrel{\rightarrow}{Q_1}$ string. The resulting color field covers the area $S_n$ of the cluster. As
$\stackrel{\rightarrow}{Q_n}=\sum^{n}_{i}\stackrel{\rightarrow}{Q_1}$, and the individual
string colors may be oriented in an arbitrary manner respective to one another,
the average $\stackrel{\rightarrow}{Q_{1i}}\stackrel{\rightarrow}{Q_{1j}}$ 
is zero, and $\stackrel{\rightarrow}{Q_n}^2 = n \stackrel{\rightarrow}{Q_1}^2$.

Knowing the charge color $\stackrel{\rightarrow}{Q_n}$, one can compute the particle spectra
produced by a single cluster of such color charge and area $S_n$ using the
Schwinger formula. For the multiplicity $\mu_n$ and average $p^2_T$ of particles,
$<p^2_T>_n$, produced by a cluster of $n$ strings one finds \cite{Ref11}-\cite{Ref12}

\begin{equation}\label{eq3}
\mu_n=\sqrt{n\frac{S_n}{S_1}}\mu_1 \quad;\quad <p^2_T>_n=\sqrt{\frac{nS_1}{S_n}}<p^2_T>_1
\end{equation}
where $\mu_1$ and $<p^2_T>_1$ are the mean multiplicity and mean $p^2_T$ of particles
produced by a single string with a transverse area $S_1=\pi r^2_0$. For
strings just touching each other $S_n=nS_1$ and hence $\mu_n=n\mu_1 ; <p^2_T>_n=<p^2_T>_1$
as expected 
(simple fragmentation of $n$ independent strings). 
In the opposite case
of maximum overlapping, $S_n=S_1$ and therefore $\mu_n=\sqrt{n}\mu_1$,$<p^2_T>_n=\sqrt{n}
<p^2_T>_1$, so that the multiplicity results maximally supressed. Notice that a certain
conservation rule holds

\begin{equation}
\frac{\mu_n}{n}<p^2_T>_n=\mu_1 <p^2_T>_1
\end{equation}
and also the scaling low

\begin{equation}\label{eq5}
<p^2_T>_n/\mu_{n}S_n=<p^2_T>_1/\mu_1 S_1
\end{equation}

In the limit of high density

\begin{equation}
<nS_1/S_n>=\frac{\eta}{1-exp(-\eta)}\equiv\frac{1}{F^2(\eta)},
\eta=N_S \pi r^2_0/\pi R^2
\end{equation}

Thus
\begin{equation}
\mu=N_S F(\eta)\mu_1\ ;\ 
<p^2_T>=<p^2_T>_1/F(\eta)
\end{equation}

The universal scaling law (\ref{eq5}) is valid for all projectiles
and targets, different energies and centralities,
being in reasonable agreement with experimental data \cite{Ref13}.
A similar scaling is found in the color glass condensate approach \cite{Ref14}.

Notice that $N_S\sim N_{coll}\sim N_A^{4/3}$ at central rapidity (at the fragmentation
region $N_S\sim N_{part}\sim N_A$). As $F(\eta)\sim N_A^{1/3} , \mu \sim N_A$.

Therefore, the multiplicity per participant does not depend on the number of participants,
there is saturation.
Numerical studies show a good agreement with SPS and RHIC data [12]. The
prediction for central $Pb-Pb$ collisions ($N_A=400$) at LHC $(\sqrt{s}=5.5\ GeV)$ is
$\mu/0.5 N_{part}=8.6$ and the total charged multiplicity per unit rapidity
(at central rapidity) is 1800. This numbers are very similar to the ones
obtained from Hera data, assuming scaling, using PQCD and the BK equation \cite{Ref16}-\cite{Ref17}.

\section{Transverse momentum fluctuations}
The behaviour of the transverse momentum fluctuations can be understood as follows:
At low densities most of the particles are produced by individual strings with
the same $<p_T>_1$, so the fluctuations must be small. Similarly, at large density,
above the percolation critical point, there is essentially one cluster formed by almost
all the strings created in the collision and therefore fluctuations are not expected either. 
Indeed, the fluctuations are expected to be maximal when the number of different
clusters becomes larger, just below the percolation critical density (see Fig. 2).
In this case in addition to the normal fluctuations around the mean transverse 
momentum of a single string, there are more fluctuations due to the different average
transverse momentum of each cluster.

Experimentally, it has been measured the quantity

\begin{equation}
F_{pt}\equiv\frac{\omega_{Data}-\omega_{random}}{\omega_{random}} \qquad
\omega=\frac{\sqrt{<p^2_T>-<p_T>^2}}{<p_T>}
\end{equation}
where $\omega_{random}$ denotes the corresponding normalized fluctuations
in the case of statistically independent particle emission.

In Fig. 3 our result \cite{Ref18} compared with the experimental data is shown.
A reasonable agreement is obtained. There is an alternative
explanation based on the occurrence at RHIC of minijets which will
enhance the $p_T$ fluctuations. At high centrality, it is well established,
the suppression of high $p_T$ particles at RHIC what can explain the suppression 
of fluctuations seen at lower centralities. According to this picture, at SPS
where the production of minijets is negligible, this behaviour is not expected,
contrary to the expectations of percolation of strings.

\begin{figure}
\resizebox{0.45\textwidth}{!}{%
  \includegraphics{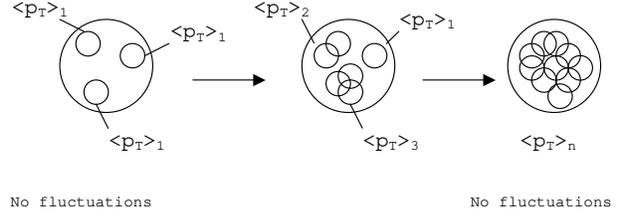}
}
\caption{Cluster formation for low, intermediate and high density respectively.}
\end{figure}

\begin{figure}
\resizebox{0.55\textwidth}{!}{%
  \includegraphics{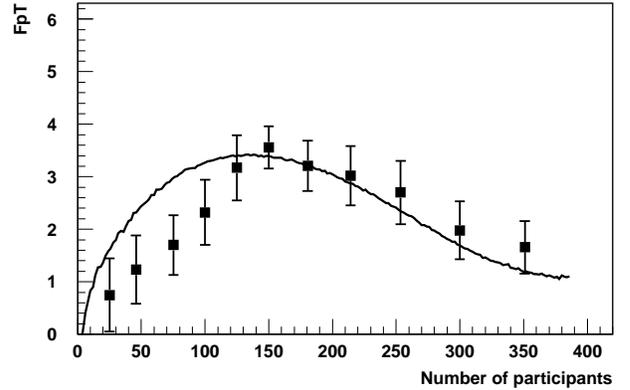}
}
\caption{$F_{p_T}(0/0)$ vs the number of participants. Experimental
data from PHENIX at $\sqrt{s}=200\ GeV$. Solid line our results.}
\end{figure}

Instead of $p_T$ fluctuations, the NA49 Collaboration \cite{Ref19} has measured
$(<n^2_->-<n_->^2)/<n_->)$ at SPS for Pb-Pb collisions, as a function
of the centrality of the collision, showing a maximum at low centrality
and being 1 at high centrality. This behaviour has nothing to do with minijets.
On the contrary, it is explained naturally in our approach \cite{Ref20}.

\section{Universal transverse momentum distributions}
In order to know the transverse momentum distributions one needs the 
fragmentation function $f(x,p_T)$ for each cluster, and the mean
squared transverse momentum distribution of the clusters, W(x), which is
related to the cluster size distribution through Eq. (\ref{eq3}). For $f(x,p_T)$ we
assume the Schwinger formula, $f(x,p_T)= exp(-p^2_T x)$, used also for
the fragmentation of a Lund string \cite{Ref21}, at first approximation $x$ is related
to the string tension or equivalently to the inverse of the mean transverse
momentum squared. For the weight function W(x) we choose the gamma distribution

\begin{equation}
W(x)=\frac{\gamma}{\Gamma(k)}(\gamma x)^{k-1} exp(-\gamma x).
\end{equation}

The reason of this choice is the following: In peripheral heavy ion
collisions there is almost not overlapping between the formed strings
and therefore the cluster size distribution is peaked around low values.
Most of the clusters are made of one single string. As the centrality increases
the number of strings grows, so there are more and more overlapping among
the strings and the cluster size distribution is strongly modified, according
to Fig. 4 where three cluster distributions corresponding to three different
centralities of the collision are shown. Each curve in Fig. 4 can be 
reproduced by gamma distributions with different $k$ values.

\begin{figure}
\resizebox{0.5\textwidth}{!}{%
  \includegraphics{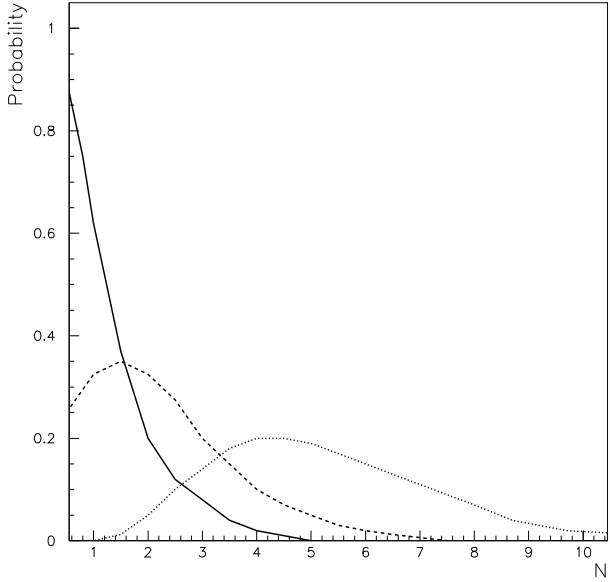}
}
\caption{Schematic representation of the number of clusters as a function
of the number of strings of each cluster at three different centralities (the
solid line corresponds to the most peripheral one and the pointed line
to the most central one).}
\end{figure}

Moreover, the increase of centrality can be seen as a transformation of the cluster
size distribution of the type

\begin{equation}
P(x)\rightarrow\frac{xP(x)}{<x>}\rightarrow\ldots\rightarrow\frac{x^kP(x)}{<x^k>}\rightarrow\ldots
\end{equation}

This kind of transformation were studied long time ago by Jona-Lasinio in connection
to the renormalization group in probabilistic theory \cite{Ref22}. Actually an increase of
the centralities is equivalent to a transformation which changes cells (single strings)
by blocks (clusters) and the corresponding variables $\mu_1$ and $<p^2_T>_1$
of the cells by $\mu_n$ and $<p^2_T>_n$. These transformations of the type of the chain (10)
have been used also to study the probability associated to events which satisfy some
requirements \cite{Ref23}.

The $\gamma$ and $k$ parameters of the gamma distribution are related to the mean $x$ and 
dispersion of the distribution through

\begin{equation}
<x>=\frac{k}{\gamma}\qquad
\frac{<x^2>-<x>^2}{<x>^2}=\frac{1}{k}
\end{equation}

We use Eq (7) to take into account the effect of overlapping of strings, and
hence $f(x,m_T)=exp(-p^2_TxF(\eta))$.  Therefore we obtain
\begin{eqnarray}\label{eq12}
\frac{1}{\left(1+\frac{F(\eta) p^2_T}{k <p^2_T>_1}\right)^k}=
\int^{\infty}_{0}dx\, exp (-p^2_T xF(\eta))\nonumber
\\\frac{\gamma}{\Gamma(k)}(\gamma x)^{k-1}exp(-\gamma x)
\end{eqnarray}
and the normalized $p_T$ distribution is

\begin{equation}\label{eq13}
f(p_T,y)=\frac{dN}{dy}\frac{(k-1) F(\eta)}{k <p^2_T>_1}
\frac{1}{\left(1+\frac{F(\eta)p^2_T}{k<p^2_T>_1}\right)^k}
\end{equation}

The equation (\ref{eq12}) can be seen as a superposition of chaotic color sources (clusters) where
$1/k$ fixes the transverse momentum fluctuations. At small density $\eta<<1$, the
strings are isolated and there are not fluctuations, $k\rightarrow\infty$. When the
density increases, there will be some overlapping of strings forming clusters, the
fluctuations increase and $k$ decreases. The minimum of $k$ will be reached when
the fluctuations in the number of strings per cluster reach its maximum. Above this
point, increasing $\eta$, these fluctuations decrease and $k$ increases. In the limit, when
only one cluster of all strings is formed, there are not fluctuations and again $k\rightarrow\infty$.

The obtained power-like behaviour $(p^2_T)^{-k}$, with an exponent $k$ related
to some intrinsic fluctuations, is common to many apparently different systems, as
sociological, biological or informatic ones. Distributions like the citations of scientific
works, or other complex networks \cite{Ref25}\cite{Ref26} where the probability $P(m)$ of having
a given node with $m$ links is described by the free scale power law $P(m)\sim (m)^{-k}$
with $k$ related to the fluctuations in the number of links obey the same behaviour. Also, it
has been shown \cite{Ref27}-\cite{Ref28} that maximization of the non extensive information Tsallis entropy
leads to the same distribution (12).

The universal behaviour indicates the importance of the common features present in those
phenomena, namely, the cluster structure and the fluctuations in the number of objets per
cluster.

>From (\ref{eq13}) one can calculate

\begin{equation}
\frac{d\,ln\,f}{d\,ln\,p_T}=\frac{-2F(\eta)}{\left(1+\frac{F(\eta)}{k}\frac{p^2_T}{<p^2_T>_i}\right)}
\frac{p^2_T>}{<p^2_T>_{1i}},
\end{equation}
where ``i'' refers to the different particle especies.

As $p^2_T\rightarrow0$ this reduces to $-2F(\eta)p^2_T/<p^2_T>_i$.

This behaviour has been confirmed by the PHOBOS Collaboration. As 
$<p^2_T>_{1p}\  \geq\  <p^2_T>_{1k}\  \geq\ <p^2_T>_{1\pi}$
the absolute value is larger for pions than for kaons and than for protons.

Now, let us discuss the interplay between low and high $p_T$. One defines
the ration $R_{CP}(p_T)$ between central and peripheral collisions as

\begin{equation}
R_{CP} (p_T)=\frac{f'(p_T,y=0)/N'_{coll}}{f(p_T,y=0)/N_{coll}}
\end{equation}
where the distribution in the numerator corresponds to higher densities $\eta'>\eta$.
In the $p_T\rightarrow0$ limit, taking into account that $\frac{2}{3}\leq\frac{k-1}{k}\leq1$ and that $F(\eta')<F(\eta)$, we obtain

\begin{equation}
R_{CP}(0)\simeq\left(\frac{F(\eta')}{F(\eta)}\right)^2<1
\end{equation}
approximately independent of $k$ and $k'$. As $\eta'/\eta$ increases, the ratio
$R_{CP}(0)$ decreases, in agreement with experimental data.

As $p_T$ increases we have

\begin{equation}
R_{CP} (p_T)\sim\frac{1+F(\eta)p^2_T/<p^2_T>_{1i}}{1+F(\eta')p^2_T/<p^2_T>_{1i}}
\end{equation}
and $R_{CP}(p_T)$ increases with $p_T$ (again, $F(\eta)>F(\eta')$)

At large $p_T$,
\begin{equation}
R_{CP} (p_T)\sim \frac{F(\eta)}{F(\eta')}\frac{k'}{k}(p^2_T)^{k-k'}
\end{equation}
which means that if we are in the low density regime $k>k'$ and $R_{CP}(p_T)>1$, and
we reproduce the Cronin effect. As we increase the energy, the density increases
and on reaches the high density regime where $k'<k$ and suppression of $p_T$ occurs. The
Cronin effect disappears at high energies and/or densities. The critical density
at which the Cronin effect disappears is the same at which the transverse
momentum fluctuations presents a maximum.

$R_{CP}(p_T)$ for two different particles, for instance $p$ and 
$\pi$, becomes, at intermediate $p_T$,

\begin{equation}
\frac{R_{CP}^p(p_T)}{R_{CP}^{\pi}(p_T)}\simeq\left(\frac{<p^2_T>_{1p}}{<p^2_T>_{1\pi}}\right)^{k'-k}
\end{equation}

As $<p^2_T>_{1p} > <p^2_T>_{1\pi}$, in the high density limit (Au-Au collisions $k'>k$)
we expect a ratio larger than 1, as the experimental data show.

As far as we approach the low density limit, the ratio decreases, becoming closer to 1 
or even lower.

A more detailed comparison with experimental data on Au-Au, d-Au collisions discussion of
the forward rapidity region can be found in reference \cite{Ref24}. An overall reasonable agreement
is obtained. It is very remarkable that such agreement is based on the universal behaviour
of the $p_T$ distribution given by Equation (\ref{eq13}).

\section{Bose-Einstein Correlations}
Most of the studies of two body Bose-Einstein (B.E) correlations have paid attention to the
parameters $R_{side}$, $R_{out}$, $R_L$ and not to the strength of the correlation, defined
by the chaoticity parameter

\begin{equation}
C_2(0,0)=\lambda
\end{equation}

Experimentally, due to Coulomb interference and to the necessary extrapolations there
are many uncertainties  in its evaluation, however, some trend of the dependence of
$\lambda$ on the multiplicity can be established. First, SPS minimum bias data for
O-C, O-Cu, O-Ag and O-Au collision show that $\lambda$ decreases
as the size of the target increases, from $\lambda=0.92$ up to $\lambda=0.16$.
However for S-Pb and Pb-Pb central collisions, where the values of $\eta$ are larger 
$\lambda$ is also larger, $\lambda\simeq0.5-0.7$.

This behaviour can easily explained in the percolation framework \cite{Ref29}. Each cluster
can be considered as a chaotic source $\lambda=1$, and the production of particles
from several clusters can be seen as the superposition of chaotic sources. In this
schemce, $\lambda=n_s/n$, being $n_s$ the number of pairs produced in the same
cluster and $n_T$ the total number of pairs. In this way, $\lambda$ is proportional to
the inverse of the number of independent sources (clusters), therefore it decreases with
the density up to the critical percolation value.
>From this critical value, it increases with density. This behaviour is shown in Fig. 5.

Similar considerations can be done concerning the s\-trength of the three body B-E correlations, $\omega$. NA44 Collaboration has obtained for S-Pb collisions, $\omega=0.20\pm0.02\pm0.19$ 
and for central Pb-Pb collisions $\omega=0.85\pm0.02\pm0.21$. STAR collaboration obtains
for central Au-Au collisions values of $\omega$ close to 1. This sharp variation from
$\omega=0.2$ to $\omega=0.8-1$ in a small range of $\eta$ is easily explained in
the framework of percolation of strings. Now $\omega$ becomes proportional to the inverse
of the squared of the number of independent sources (clusters) what can acomodate stronger
variation compared to the case of two body. In Fig. 6 our result is shown \cite{Ref30}.

\begin{figure}
\resizebox{0.5\textwidth}{!}{%
  \includegraphics{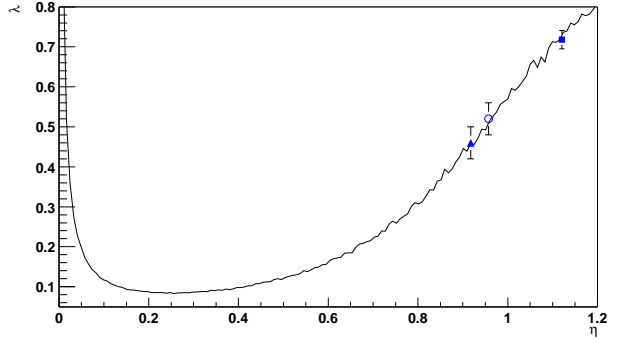}
}
\caption{$\lambda$ as a function of $\eta$. Experimental points are for semicentral S-Pb
collisions (filled triangle), 18\% central Pb-Pb collisions (non filled circle) and 10\% 
central Pb-Pb collisions (filled box) at SPS.}
\end{figure}

\begin{figure}
\resizebox{0.5\textwidth}{!}{%
  \includegraphics{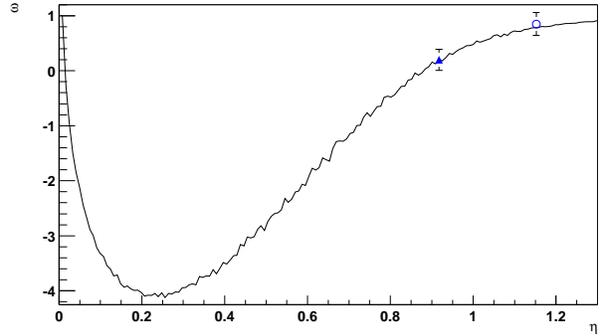}
}
\caption{$\omega$ as a function of $\eta$. Experimental points are for S-Pb semicentral
collisions and 9\% central Pb-Pb collisions at SPS.}
\end{figure}

\section{Forward-Backward Correlations} 
An useful observable to check the percolation
approach is the forward-backward correlation measured by the quantity

\begin{equation}
D_{FW}=<n_Fn_B>-<n_F><n_B>
\end{equation}
where $n_{F(B)}$ denotes the multiplicity in a forward (backward) rapidity interval. In 
order to eliminate the short range correlations, the forward and backward o intervals should
be separated by at least one unit of rapidity. On general grounds, one can see that
$D_{FW}$ is proportional to the fluctuations on the number of independent sources, (or
clusters in our case)\cite{Ref31}-\cite{Ref32}. At very low density, $D_{FW}$ should be very small, increasing with the
density up to a maximum related to the largest number of clusters. At very high density,
there is essentially only are cluster and hence $D_{FW}$ becomes small again.

There are some experimental data measuring the parameter b, through

\begin{equation}
<\mu_B>_F=a+b\mu_F
\end{equation}
where $b\equiv D_{FB}/D{FF}$. The data on $pp$ and $pA$ show on increase of $b$
with energy and density. Our prediction for high density is that $b$ will decrease.
Measurements of $D_{FW}$ or $b$ as $a$ function of centrality would be welcome.

\section{Conclusion}
The percolation of partons and strings can describe rightly several observables, namely
$J/\psi$ suppression, multiplicities, transverse momentum fluctuations, transverse
momentum distributions and B-E correlations. The behaviour of all of them
has a common physical basis: the clustering of color sources and the dependence of the number
of cluster on the density. In this way, the threshold of $J/\psi$ suppression, the maximum
of transverse momentum fluctuations, the suppression of the Cronin effect and the turnover of the
dependence of the strength of two and three body correlation with the energy are related
to each other and all of them point out a percolation phase transition. Another test
of this transition is the measurements of forward-backward correlations and also the
multiplicity distributions not discussed here \cite{Ref33}.

Many of the results obtained in the framework of percolation of strings are very similar to the
one obtained in the color glass condensate (CGC). In particular, very similar scaling
lows are obtained for the product and the ratio of the multiplicities and transverse
momentum. For this reason, it is very tempting to identify the momentum $Q_s$ which
established the scale in CGC with the corresponding are in percolation of string. In this way

\begin{equation}\label{eq23}
Q^2_s=\frac{k<p^2_T>_1}{F(\eta)}
\end{equation}
The consequences of Eq. (\ref{eq23}) are under study.

Acknowledgements. I thank the organizers for such a nice meeting and E.G. Ferreiro for a
critical reading of the manuscript.
This work has been done under contracts FPA2002-01161 of CICyT of Spain, and
PGIDIT03PXIC-20612PN from Galicia.

%

%
%
%
%

\end{document}